\documentclass[12pt]{iopart}
\usepackage{iopams}
\usepackage{txfonts}
\usepackage{graphicx}

\def\arcmin{\hbox{$^\prime$}}

\def\utw{\smash{\rlap{\lower5pt\hbox{$\sim$}}}}
\def\udtw{\smash{\rlap{\lower6pt\hbox{$\approx$}}}}

\def\farcs{\hbox{$.\!\!^{\prime\prime}$}}

\begin{document}
\article[Short title]{Star-forming Dwarf Galaxies: Ariadne's Thread in the
  Cosmic Labyrinth}{Dwarf galaxies and the Magnetisation of the IGM}

\author{Uli Klein$^1$, P. Papaderos$^2$}

\address{$^1$Argelander-Institut f\"ur Astronomie, Auf dem H\"ugel 71, D-53501 Bonn, Germany}
\address{$^2$Instituto de Astrofísica de Andalucía, Camino Bajo de Huétor, 50, 18008 Granada, Spain}
\ead{\mailto{uklein@astro.uni-bonn.de}}

\begin{abstract}
With the operation of LOFAR, a great opportunity exists to shed light on
a problem of some cosmological significance. Diffuse radio synchrotron 
emission not associated to any obvious discrete sources as well as 
Faraday rotation in clusters of galaxies both indicate that the 
intergalactic or intracluster medium (IGM, ICM) is pervaded by a 
weak magnetic field, along with a population of relativistic particles. 
Both, particles and fields must have been injected into the IGM either 
by Active Galactic Nuclei (AGN) or by normal star-forming galaxies. 
Excellent candidates for the latter are starburst dwarf galaxies, which 
in the framework of hierarchical structure formation must have been around 
in large numbers. If this is true, one should be able to detect extended 
synchrotron halos of formerly highly relativistic particles around local 
starburst or post-starburst dwarf galaxies. With LOFAR, one should easily 
find these out to the Coma Cluster and beyond. 
\end{abstract}



\section{Introduction}

Dwarf galaxies play a key role in the enrichment of the ICM or IGM, not 
only as far as heavy elements are concerned, but possibly also regarding 
the magnetisation. According to the standard bottom-up scenario of galaxy 
formation, primeval galaxies must have injected much of their (enriched) 
interstellar medium (ISM) into the IGM during the initial bursts of star 
formation, thereby ''polluting'' large volumes of intergalactic space 
because of their high number density. By the same token, the following 
two properties render dwarf galaxies potentially very efficient in 
injecting a relativistic plasma into their surroundings: first, they 
exist in large numbers, and second, they possess low escape velocities, 
making it easier to expel their interstellar gas, as compared to massive 
spiral galaxies. Such galactic winds are in fact seen in some prototypical 
low-mass galaxies in the local universe. 

\begin{figure}[h]
\begin{center}
 \includegraphics[width=6.0in]{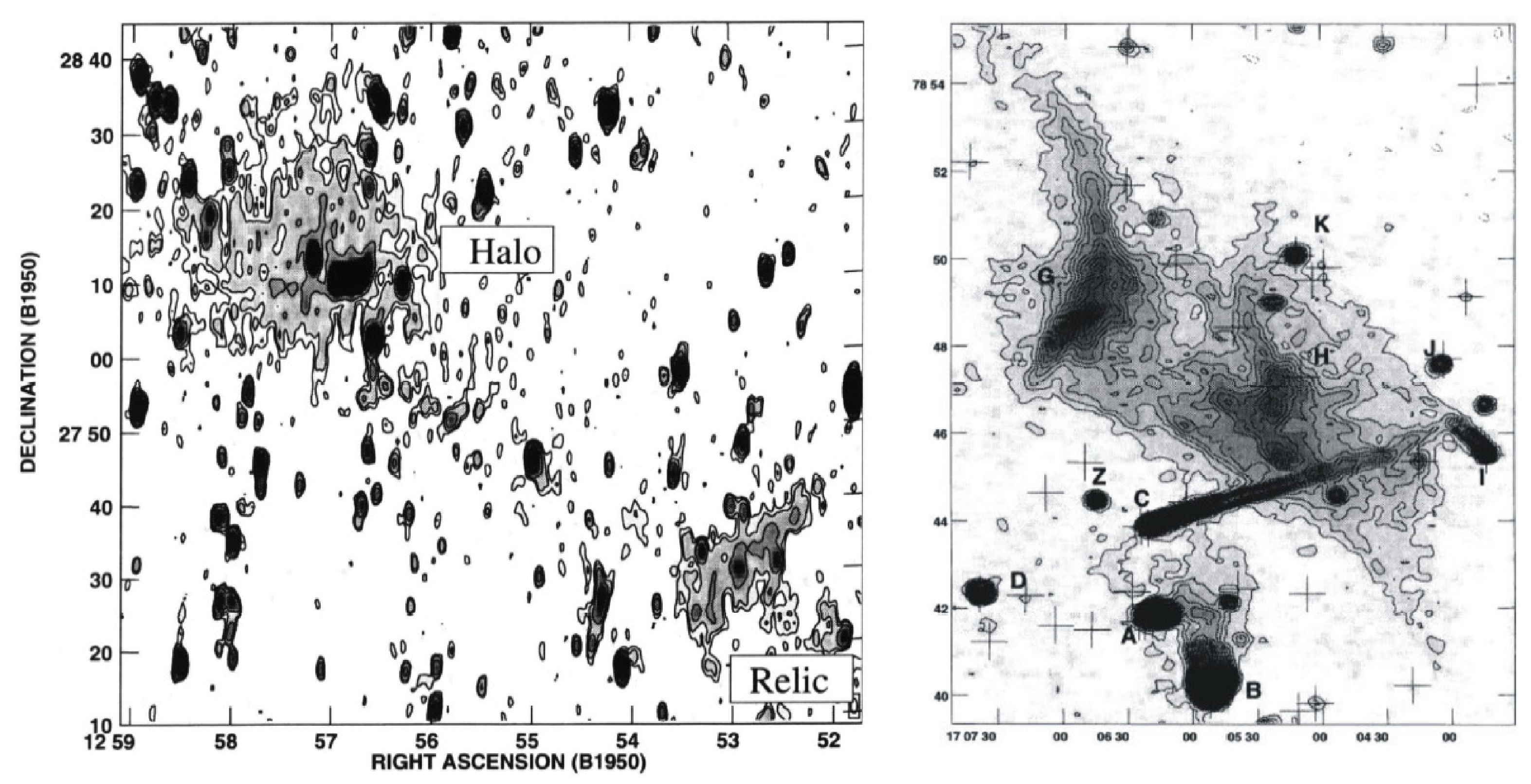} 
 \caption{Diffuse radio continuum emission from clusters of galaxies: central 
radio halo and peripheral structure in the Coma Cluster (left, from 
\cite{Giovannini_etal93}), and radio relic in A\,2556 (right, from 
\cite{Roettgering_etal94}.)}
\label{fig1}
\end{center}
\end{figure}

\section{Outflows and the magnetisation of galaxy clusters}

Galaxy clusters are known to be pervaded by a relativistic plasma, 
i.e. particles with (mildly) relativistic energy and magnetic fields. 
In the centres of some rich clusters, radio halos have been found, 
produced by either primary or secondary electrons (see Fig.~\ref{fig1}, 
left). The problem with primary electrons is their relatively short 
lifetime, limited by synchrotron and inverse-Compton losses: 
\begin{equation}
\label{eqn1}
t_{1/2}	= 1.59 \cdot 10^9 \cdot \frac{B^{1/2}}{B^2 + B_{cmb}^2} 
\left [ \left (\frac{\nu}{\rm GHz} \right ) (1 + z) \right]^{-1/2}
\,\, \rm yr \;,
\end{equation}
where
\begin{equation}
B_{cmb} = 3.25 \, (1 + z)^2 \,\, \rm \mu G 
\end{equation}
is the equivalent magnetic field strength of the cosmic microwave 
background. This latter equation results from the similar dependences 
of synchrotron and inverse-Compton losses, 
\begin{equation}
\dot{E}_{synch} \propto - u_{mag} \cdot E^2 
\end{equation}
\begin{equation}
\dot{E}_{IC} \propto - u_{rad} \cdot E^2 \, ,
\end{equation}
where $u_{mag}$ is the magnetic energy density $u_{mag}	= 
B^2/8 \pi$, and $u_{rad}$ is the energy density of the radiation field (which
for the perfect black-body CMB radiation is $\propto T_{cmb}^4$). 
Eqn.~(\ref{eqn1}) then tells us that, for instance, with a magnetic 
field of $B = 1 ~ \mu$G, particles radiating at 1.4~GHz will be 
rendered invisible after $t_{1/2} = 10^8$~yr in the local universe 
($z = 0$). Hence, primary electrons require continuous injection. 
However, diffuse radio halos are never associated with any obvious 
``fresh'' sources, one therefore has to invoke secondary electrons. 
These are produced by hadronic collisions of relativistic protons 
with the thermal gas in a pion-muon chain. Owing to their much larger 
mass, protons have synchrotron lifetimes exceeding a Hubble time. 
Finally, it should pointed out that completely independent evidence 
for the magnetisation of (at least the) central regions of galaxy 
clusters comes from the observation of Faraday rotation (e.g. 
\cite{Clarke_etal01}). 

Clusters undergoing large-scale merging frequently exhibit so-called 
radio relics, mostly on their periphery (Fig.~\ref{fig1}, right). In 
contrast to radio halos, these are significantly polarised, probably 
reflecting magnetic-field enhancement in the compression zones where 
the subclusters produce large-scale shockwaves during their mutual 
penetration. Naturally, this also provides an efficient acceleration 
mechanism. Particles that were formerly highly relativistic (electrons 
in particular) have cooled via synchrotron and inverse-Compton losses. 
During a cluster merger, they are boosted to high energies again, and 
the acceleration regions shine up in the radio regime.

Kronberg et al. (\cite{Kronberg_etal99}) were the first 
to raise the question whether low-mass galaxies could have made a 
significant contribution to the magnetisation of the IGM (apart from 
more massive starburst galaxies and AGN). Owing to their large number 
(observed and predicted in a CDM cosmology) and their injection of 
relativistic particles as described above, they could have played a 
cardinal role in the context of this cosmologically important scenario. 
If true, it is to be expected that dwarf galaxies are ``wrapped'' in 
large envelopes of previously highly relativistic particles - and 
magnetic fields, which are pushed out of them during epochs of 
vigorous star formation. \cite{Bertone_etal06} 
have discussed this more quantitatively and made predictions for the 
strengths of magnetic seed fields, to be then amplified by large-scale 
dynamos over cosmic time. In particular, they also predict the existence 
of magnetic voids. 

Of course, in magnetising the ICM/IGM, low-mass galaxies have been 
competing with AGN. Judging from the radio luminosities of the 
``culprits'', it is clear that nevertheless low-mass galaxies may 
have contributed significantly. A typical starburst dwarf galaxy emits 
a monochromatic radio luminosity of $P_{\rm 1.4~GHz} \approx 
10^{20.5}$~W~Hz$^{-1}$, while this figure is $P_{\rm 1.4~GHz} \approx 
10^{24.7}$~W~Hz$^{-1}$ for radio galaxies in the FRI/II transition 
regime. Hence, the radio power produced by AGN is some 15000 times 
larger than that of dwarf galaxies. However, ($\Lambda$CDM) cosmology 
helps at this point, since dwarf galaxies have come in huge numbers 
if the bottom-up scenario of structure formation holds. Furthermore, 
the lifetime of radio galaxies is limited (\cite{Bird_etal08}), 
$\tau_{life} \approx 1.5 \cdot 10^7$~yr, with 
duty cycles of $\tau_{life} \approx 8 \cdot 10^8$~yr. Hence, this 
yields an effective activity period of $\tau_{active} \approx 2 \cdot 
10^8$~yr~$\approx 0.015 \times \tau_{\rm Hubble}$. 

Finally, it should be noted that central so-called ``mini-halos'', 
which are bright extended radio sources located in the centres of 
cooling-flow clusters (Perseus\,A, Hydra\,A, Virgo\,A) cannot do the 
magnetisation job: they are pressure-confined.

\section{Local templates}

\begin{figure}[h]
\begin{center}
 \includegraphics[width=5.0in]{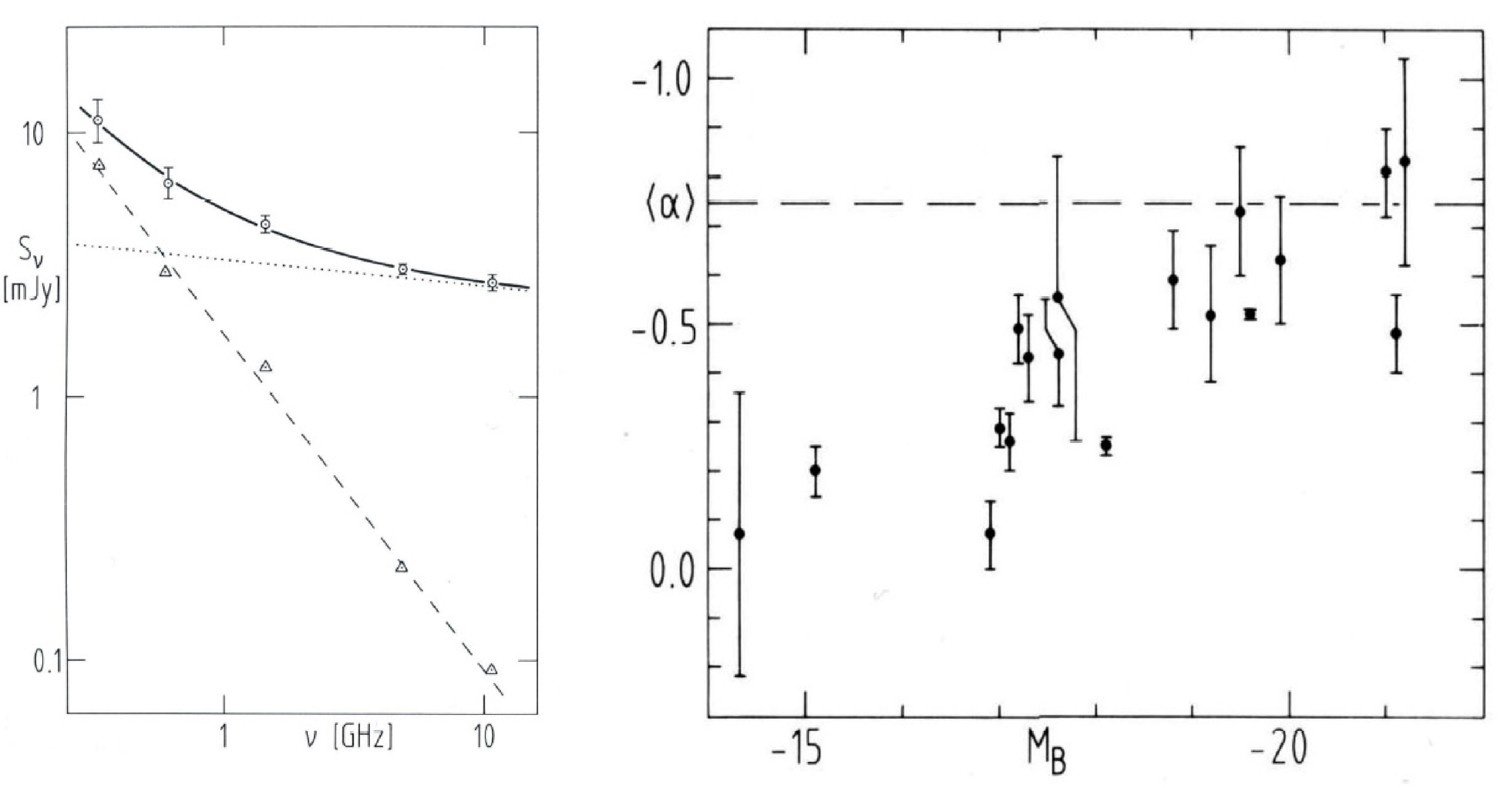} 
 \caption{Left: radio continuum spectrum of the BCDG II\,Zw\,70, with 
the thermal (free-free, dotted line) and the nonthermal (synchrotron, 
dashed line) components indicated. The solid line represents the total 
flux density fitted to the measurements (from \cite{Skillman88}). Right: spectral indices of the integrated radio 
emission of a sample of low-mass galaxies, mostly BCDG (from 
\cite{Klein_etal91}).}
   \label{fig2}
\end{center}
\end{figure}

Measurements of the radio continuum radiation of dwarf galaxies over a 
large frequency range have shown that ongoing star formation in them
is accompanied by enhanced radio continuum emission (\cite{Klein82}, 
\cite{Klein_etal91}). Owing to their shallow gravitational potentials, 
the containment of relativistic cosmic-ray particles in such galaxies 
is, however, low, as inferred from studies of the spectral index and 
magnetic-field structure. Fig.~\ref{fig2} demonstrates this: in contrast 
to massive spiral galaxies, whose radio continuum is dominated by 
synchrotron radiation at cm wavelengths (\cite{Gioia_etal82}) there is 
a lack of it for dwarf galaxies. The lack is more pronounced the lower 
the luminosity (hence mass) of the dwarf galaxy (\cite{Klein_etal91}). 
These relativistic particles streaming into the halos of dwarf galaxies 
lose their energy on a time scale of $\sim 10^8$~yr according to 
Eqn.~(\ref{eqn1}) via synchrotron and Inverse-Compton radiation. They 
are thus quickly rendered invisible at cm wavelengths, while their 
synchrotron emission will still be detectable at metre waves where 
their lifetime is 5 to 10 times longer. 

The containment of the relativistic plasma and its possible transport 
out of a galaxy is obviously influenced by the overall magnetic-field 
configuration. In massive disk galaxies (spirals) the face-on view 
shows that the magnetic field closely follows the spiral arms, while 
the edge-on view discloses a plane-parallel magnetic field close to the 
galactic planes, while the halo field flares at large galacto-centric 
distances, rendering its overall appearance 'X-shaped' (\cite{Beck08}).

The existence of winds in low-mass galaxies is inferred from the observed 
kinematics of the gas (measured with slit spectroscopy), but can arguably 
be also inferred from measurements of the temperature of the hot 
(X-ray-emitting) gas. For instance, \cite{Martin98} found the outflow 
velocities in NGC\,1569 to exceed the escape speed, and 
\cite{dellaCeca_etal96} derived a temperature of its hot, 
X-ray-emitting gas to exceed the virial temperature. 

\begin{figure}[h]
\begin{center}
 \includegraphics[width=6.0in]{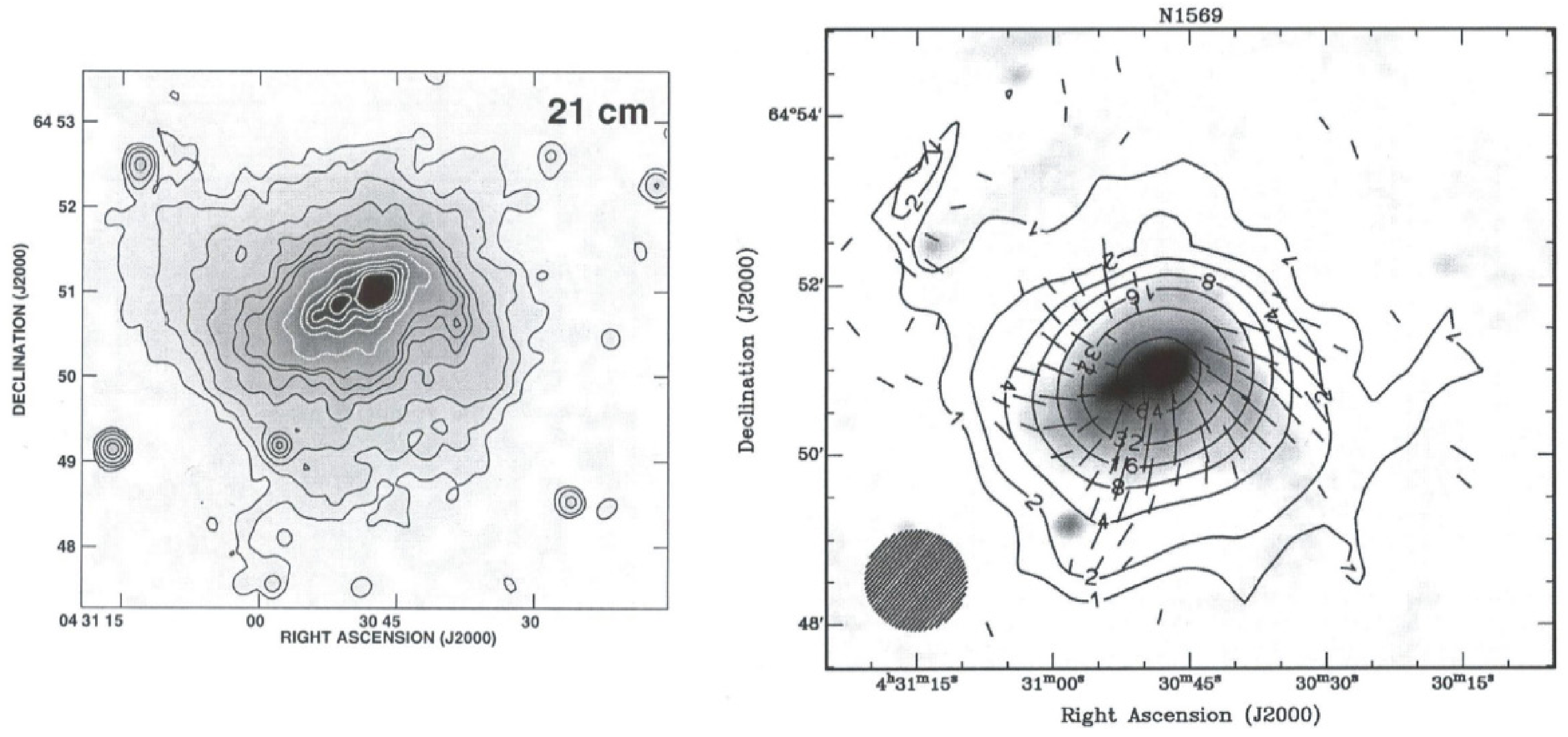} 
 \caption{Synchrotron radiation of the starburst dwarf galaxy NGC\,1569 
(from \cite{Kepley_etal08}). 
A radio halo is visible at 21~cm wavelength (VLA observation, left), and 
the magnetic field orientation, derived from the linear polarisation at 
2.8~cm wavelength, is primarily oriented radially outward (measurements 
with the Effelsberg 100-m telescope, right).}
   \label{fig3}
\end{center}
\end{figure}

The transport of a relativistic plasma out of this galaxy is strongly 
suggested by two observations (\cite{Kepley_etal08}). First, 
NGC\,1569 possesses a radio halo, extending out to about 2~kpc at 1.4~GHz 
(see Fig.~\ref{fig3}, left). Second, the projected orientation of its 
magnetic field as deduced from measurements of the linear radio 
polarisation is radial throughout (Fig.~\ref{fig3}, right). This 
magnetic field is just dragged along with the wind, as the energy 
density of the wind is about 30 times that of the magnetic field. A 
similar finding was made by \cite{Chyzy_etal00} 
for the dwarf irregular NGC\,4449, which also possesses a low-frequency 
radio halo (\cite{Klein_etal96}).

\section{Lifetime and size of radio halos}

\begin{figure}[h]
\begin{center}
 \includegraphics[width=3.4in]{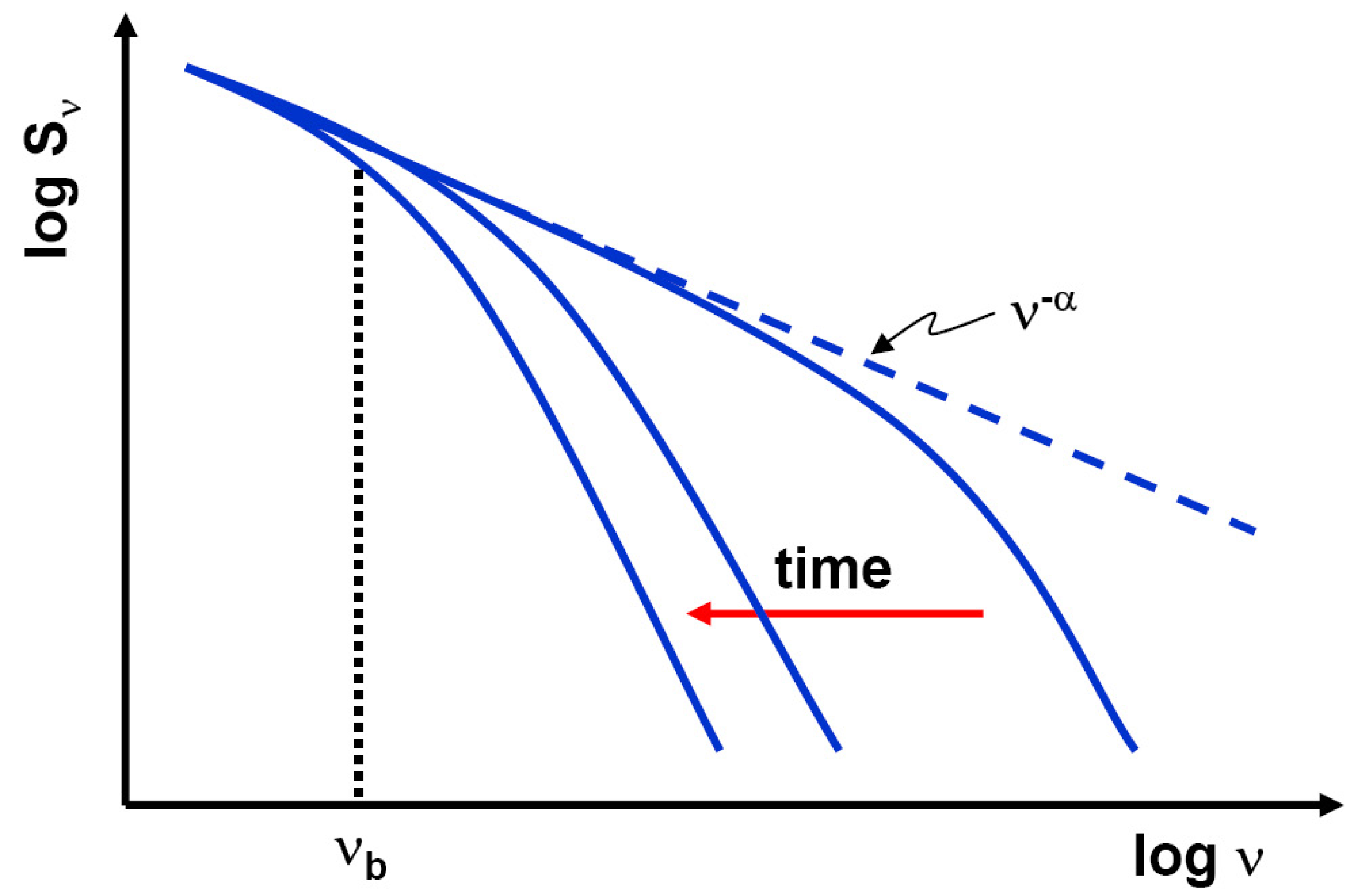} 
 \caption{Temporal evolution of synchrotron spectra in galaxies.}
   \label{fig4}
\end{center}
\end{figure}

The lifetime of low-frequency halos around dwarf galaxies can be 
estimated using Eqn.~(\ref{eqn1}). With $\mu$G magnetic fields in 
their surroundings the dominant loss mechanism will be inverse 
Compton, the lifetime of relativistic electrons radiating at 120~MHz 
then being of order 500~Myr, much longer than for those seen at cm 
wavelengths. This is also illustrated in Fig.~\ref{fig4}, which 
sketches the temporal variation of an ``aging'' synchrotron spectrum. 
Assume that the energy spectrum of the particles initially extended 
to infinity. As time elapses, high-energy particles quickly lose 
their energy, producing a break in the spectrum. This gives rise 
to a corresponding break in the radiation spectrum, which wanders 
more and more slowly towards lower frequencies as time elapses. 
Hence, it takes much more time for this break to arrive at 
frequencies observed, e.g., with LOFAR. The break frequency 
measures the time elapsed since the last starburst with its 
supernova activity has ceased. 

While aging, particles that have left a dwarf galaxy by a wind 
will slowly diffuse away from it. This can happen with a maxmimum 
speed corresponding to the Alfvenic one, i.e. 
\begin{equation}
v_A = 2.2 \cdot \left ( \frac{B}{\mu \rm G} \right ) 
          \cdot \left ( \frac{n_e}{\rm cm^{-3}}	\right )^{-\frac{1}{2}}  
\; \rm km~s^{-1}, 
\end{equation}
where $n_e$ is the number density of thermal electrons (or protons) 
of the surrounding medium. Taking $B = 1~\mu$G and $n_e = 
0.001$~cm$^{-3}$, the relativistic particles radiating at 120~MHz 
could move out to $30 \ldots 40$~kpc within 500~Myr, or correspondingly 
further when caught at still lower frequencies. Hence, such halos could 
have total sizes of $60 \ldots 80$~kpc. 

\section{Observations with LOFAR}

LOFAR will easily detect low-frequency halos of dwarf galaxies out 
to large (100~Mpc, Coma Cluster) distances. The LOFAR survey, which 
will cover about half the sky and will be conducted within a year's 
time, will provide a 5-$\sigma$ flux limit of 0.1~mJy/b.a. at 120~MHz 
in 1 hour (with a synthesized beam of 1\farcs3, for long baselines). 
Using the synchrotron halo of NGC\,4449 as a template, we can estimate 
the expected brightness at LOFAR frequencies, which is about 0.4~mJy/b.a. 
at 120~MHz. Its detection should hence be feasible. Assuming a 60~kpc 
diameter as estimated above and placing it at $D \approx 100$~Mpc  
it would be seen with a $\sim$~2\arcmin\ angular extent. In a cluster 
or group environment, such halos are likely to trail behind the galaxy 
as it moves in the cluster potential with speeds much larger than 
the diffusion speed of the relativistic particles. This will actually 
provide some information on the proper motion of dwarf galaxies 
having undergone star formation over the past $10^9$~yrs. An impressive 
example of such trailing low-frequency radio structures is seen in the 
Perseus Cluster, where the head-tail radio source NGC\,1265 leaves 
behind a huge radio tail that can be traced over a projected path of 
1.1~Mpc at low frequencies (\cite{Sijbring98}) 

So, low-frequency radio emission ``memorises'' starburst activity up to
about 1~Gyr after its termination, a time scale much longer than that 
of any other integral property of galaxies (H$\alpha$ and FIR luminosity, 
or broad-band optical/UV/IR colours). This emission therefore also provides 
a powerful diagnostic tool to search for post-starburst galaxies and to 
explore their recent-to-past star-formation rate (SFR) and magnetic field 
evolution. 

In this context, LOFAR will also deliver important new information 
to address the long-standing question pertaining to possible 
evolutionary links between the two main classes of late-type 
dwarf galaxies, dwarf irregulars (dIs) and blue compact dwarf 
galaxies (BCDGs). BCDGs in a post-starburst phase will be easily 
identifiable by their weak radio emission at cm wavelengths 
in conjunction with their bright radio halos at meter wavelengths. 
By contrast, non-starbursting late-type dwarf galaxies will be 
rendered undetectable at cm wavelengths, while their low-frequency 
radio continuum emission should still exhibit the synchrotron halos 
created during past activity phases. Hence, if the standard 
evolutionary scenario, i.e. from BCDGs to dIs, is correct, then 
LOFAR will discover a large population of quiescent dIs with 
low-frequency radio halos.

\ack
UK expresses his deep gratitude for the warmhearted hospitality he
experienced at {\it The Orthodox Academy of Crete} and, in particular,
to Prof. Alexandros K. Papaderos and his family.

\end{document}